\pdfoutput=1

\documentclass[prl,aps,twocolumn,showpacs,preprintnumbers,amssymb,nofootinbib,raggedbottom,10pt]{revtex4-1}
\usepackage{hyperref,graphicx,array,multirow,color,amsmath,mathrsfs,titlesec}
\titleformat{\section}{\centering\normalsize\normalfont\bf}{\thesection}{1em}{}

\hypersetup{pdftitle={},pdfcreator={},linkcolor=[rgb]{0.15,0.35,0.75},colorlinks=true,citecolor=[rgb]{0.675,0,0.2},urlcolor=[rgb]{0.15,0.35,0.65}}

\newcommand{\fwbox}[2]{\text{\makebox[#1][c]{$\hspace{-150pt}\displaystyle#2\hspace{-150pt}$}}}
\newcommand{\fwboxL}[2]{\text{\makebox[#1][l]{$#2$}}}
\newcommand{\fwboxR}[2]{\text{\makebox[#1][r]{$#2$}}}
\renewcommand{\phi}{\varphi}

\newcommand{\eq}[1]{\vspace{-3.5pt}\begin{equation}\hspace{2pt}#1\hspace{-0pt}\vspace{-3.5pt}\end{equation}}

\newcommand{\fig}[2]{\vcenter{\includegraphics[scale=#1]{#2}}}
\newcommand{\mi}{\raisebox{0.75pt}{\scalebox{0.75}{$\hspace{-1pt}\,-\,\hspace{-0.75pt}$}}}
\renewcommand{\pl}{\raisebox{0.75pt}{\scalebox{0.75}{$\hspace{-1pt}\,+\,\hspace{-0.75pt}$}}}
\newcommand{\proj}[1]{\raisebox{0.75pt}{\big[}\hspace{-0.75pt}#1\hspace{-0.75pt}\raisebox{0.75pt}{\big]}}

\newcommand{\equivR}{\fwbox{14.5pt}{\hspace{-0pt}\fwboxR{0pt}{\raisebox{0.47pt}{\hspace{1.25pt}:\hspace{-4pt}}}=\fwboxL{0pt}{}}}
\newcommand{\equivL}{\fwbox{14.5pt}{\fwboxR{0pt}{}=\fwboxL{0pt}{\raisebox{0.47pt}{\hspace{-4pt}:\hspace{1.25pt}}}}}

\newcommand{\F}{\mathfrak{F}}
\newcommand{\U}{\mathfrak{U}}
\newcommand{\T}{\mathfrak{T}}

\makeatletter\DeclareRobustCommand*{\bfseries}{\not@math@alphabet\bfseries\mathbf\fontseries\bfdefault\selectfont\boldmath}\makeatother

\begin{document}
\title{\texorpdfstring{A (Bounded) Bestiary of Feynman Integral Calabi-Yau Geometries\\[-18pt]~}{A (Bounded) Bestiary of Feynman Integral Calabi-Yau Geometries}}
\author{Jacob~L.~Bourjaily}
\affiliation{Niels Bohr International Academy and Discovery Center, Niels Bohr Institute,\\University of Copenhagen, Blegdamsvej 17, DK-2100, Copenhagen \O, Denmark}
\author{Andrew~J.~McLeod}
\affiliation{Niels Bohr International Academy and Discovery Center, Niels Bohr Institute,\\University of Copenhagen, Blegdamsvej 17, DK-2100, Copenhagen \O, Denmark}
\author{Matt~von~Hippel}
\affiliation{Niels Bohr International Academy and Discovery Center, Niels Bohr Institute,\\University of Copenhagen, Blegdamsvej 17, DK-2100, Copenhagen \O, Denmark}
\author{Matthias~Wilhelm}
\affiliation{Niels Bohr International Academy and Discovery Center, Niels Bohr Institute,\\University of Copenhagen, Blegdamsvej 17, DK-2100, Copenhagen \O, Denmark}

\begin{abstract}
We define the {\it rigidity} of a Feynman integral to be the smallest dimension over which it is non-polylogarithmic. We prove that massless Feynman integrals in four dimensions have a rigidity bounded by $2(L\mi1)$ at $L$ loops provided they are in the class that we call {\it marginal}: those with $(L\pl1)D/2$ propagators in (even) $D$ dimensions. We show that marginal Feynman integrals in $D$ dimensions generically involve Calabi-Yau geometries, and we give examples of finite four-dimensional Feynman integrals in massless $\phi^4$ theory that saturate our predicted bound in rigidity at all loops.
\end{abstract}
\maketitle

\section{Introduction}\label{introduction_section}\vspace{-14pt}

One-loop Feynman integrals in generic quantum field theories are known to be polylogarithmic (see e.g.\ \mbox{refs.\ \cite{Davydychev:1997wa,Arkani-Hamed:2017ahv}}). 
More generally, multiple polylogarithms have been found to suffice for sufficiently simple quantities, and at sufficiently low multiplicity or low loop orders (see e.g.\ \mbox{refs.\ \cite{Beisert:2006ez,CaronHuot:2011ky,Bourjaily:2015bpz,Bourjaily:2016evz,Henn:2016jdu,DelDuca:2016lad,Caron-Huot:2016owq,Dixon:2016nkn,Almelid:2017qju}}). Nevertheless, it has been known for some years that Feynman integrals of worse-than-polylogarithmic complexity are relevant to quantum field theories. Integrals with arbitrarily worse complexity were first observed in massive theories in two dimensions (see e.g.\ \mbox{refs.\ \cite{Broadhurst:1993mw,Bloch:2013tra,Bloch:2016izu,Bloch:2014qca,Broadhurst:2016myo,Bogner:2017vim}}), but are now known to be important even in massless, integrable theories in four dimensions (see e.g.\ \mbox{refs.\ \cite{CaronHuot:2012ab,ArkaniHamed:2012nw,Bourjaily:2015jna,Bourjaily:2017wjl,Bourjaily:2017bsb,Bourjaily:2018ycu}}).

We define the geometric `rigidity' of a Feynman integral to be its degree of `non-polylogarithmicity'. More concretely, after eliminating a maximal number of rational integrations we imagine all the ways in which a Feynman integral can be expressed as a sum of polylogarithms of weight \mbox{$\geq\!w$} integrated over a space of higher genus or dimension than a Riemann sphere; maximizing the weight $w$ minimizes the dimension of the space that remains---which we use to define the {rigidity} of the integral. (Some integrals will consist of a sum of terms with different rigidity; we consider the rigidity of the integral to be the maximum rigidity that contributes.) Thus, polylogarithms have rigidity 0, while the (two-dimensional) massive sunrise integral~\cite{Broadhurst:1993mw,Caffo:1998du,Laporta:2004rb,Kniehl:2005bc,Groote:2005ay,Groote:2012pa,Bailey:2008ib,MullerStach:2011ru,Adams:2013kgc,Bloch:2013tra,Remiddi:2013joa,Adams:2014vja,Adams:2015ydq,Bloch:2016izu,Adams:2017ejb,Remiddi:2016gno}, and the (four-dimensional) kite~\cite{Adams:2016xah,Remiddi:2016gno,Adams:2017ejb,Bogner:2017vim} and massless double-box integrals~\cite{CaronHuot:2012ab,Nandan:2013ip,Paulos:2012nu,Bourjaily:2017bsb} have rigidity 1 (as they involve integration over a one-dimensional variety with genus one). Feynman integrals with higher rigidity are also known to exist:  the massive $L$-loop banana integral in two dimensions (see e.g.\ \mbox{refs.\ \cite{Bloch:2014qca,Bloch:2016izu,broadhurstprivate,mirrors_and_sunsets,Groote:2005ay}}) and the massless $L$-loop traintrack integral in four dimensions~\cite{Bourjaily:2018ycu} both have rigidity $L\mi1$.

In this Letter, we probe the limits of Feynman-integral rigidity. In particular, we prove that a large class of massless Feynman integrals in four dimensions have a rigidity bounded by $2(L\mi1)$ at $L$ loops, and we provide explicit examples that saturate this bound. Maximal rigidity is easiest to demonstrate for the case of integrals that we call {\it marginal}: those $L$-loop integrals involving exactly $(L\pl1)D/2$ propagators in (even) $D$ dimensions. Moreover, we show using the Symanzik polynomial formalism that marginal integrals generically involve Calabi-Yau geometries with dimension equal to the rigidity of the integral.  Thus, we describe examples of massless integrals in four dimensions that involve a K3 surface at two loops, CY${}_4$'s at three loops, CY${}_6$'s at four loops, and so on---all examples exceeding previously known limits of rigidity. Our searches have uncovered a veritable bestiary of examples which saturate our predicted bound, some to all loop orders.

All of our results are described in terms of the Symanzik polynomial representation of Feynman integrals. We review this formalism momentarily, and use this to motivate the notion of {\it marginality} described above---including its relevance to (proving) a Calabi-Yau condition to be qualified below. We show that finite marginal integrals in two dimensions have rigidity \mbox{$\leq\!(L\mi1)$}, and that this bound is saturated if (and only if) all propagators are massive---thus reproducing the well-known fact that the $L$-loop massive banana integral can be expressed as a logarithm integrated over a $(L\mi1)$-dimensional (singular) Calabi-Yau manifold \cite{Bloch:2014qca,Bloch:2016izu,broadhurstprivate,mirrors_and_sunsets}. We then generalize this to four dimensions and show that finite marginal integrals involving massless propagators have a bounded rigidity---and we describe a number of examples which saturate this bound. We conclude with some general observations (and wild conjectures), and discuss how the examples here compare with the planar limit.

\vspace{-18pt}\section{Review: Symanzik Form of Loop Integrals}\vspace{-14pt}\label{sec:symanzik}

The examples in which we are presently most interested are easiest to understand using the Symanzik polynomial formalism. (We refer the more interested reader to e.g.\ \mbox{ref.\ \cite{Smirnov:2004ym}} for a more thorough discussion.) Let us restrict our attention to scalar Feynman integrals (those with loop-independent numerators) involving propagators with unit powers. Associating a Schwinger (or more precisely, an `$\alpha$') parameter $x_i$ to the $i$th of $E$ propagators and integrating out each of the ($L D$)-dimensional loop momenta results in the following `Symanzik' representation of a loop integral:
\vspace{-2pt}\eq{I=\int_{x_i\geq0}\hspace{-17.5pt}\proj{d^{E-1}x_i}\frac{\U^{E-(L+1)D/2}}{\F^{E-L D/2}}\,.\label{symanzik_rep}\vspace{-2pt}} 
Here, $\proj{d^{E-1}x_i}$ is the integration measure over $E$ homogeneous variables $x_i$ on $\mathbb{P}^{E-1}$, where we have dropped some conventionally included numerical prefactors. The first and second Symanzik polynomials, $\U$ and $\F$ in (\ref{symanzik_rep}), are defined in terms of the Feynman graph. Specifically, letting $e_i$ denote the $i$th edge, 
\vspace{-0pt}\eq{\begin{split}\U\equivR&\sum_{\{T\}\in\T_1}\!\!\raisebox{-0pt}{$\left(\rule{0pt}{10.5pt}\right.$}\!\prod_{e_i\notin T}x_i\!\raisebox{-0pt}{$\left.\rule{0pt}{10.5pt}\right)$}\,;\\
\F\equivR&\Bigg[\sum_{{\{T_1,T_2\}\in\T_2}}\hspace{-15pt}s_{T_1}\Big(\hspace{-10pt}\prod_{{\hspace{5pt}e_i\notin T_1\cup T_2\hspace{-5pt}}}\hspace{-10pt}x_i\Big)\Bigg]+\U\sum_{e_i}x_im_i^2\,.\\[-6pt]\end{split}\label{symanzik_polys_defined}\vspace{-0pt}}
Here, $m_i$ denotes the mass of the $i$th propagator, $\T_k$ denotes the set of spanning $k$-forests of the graph,\footnote{A spanning $k$-forest of a graph is a subgraph involving all vertices and having $k$ connected components, each of which are trees.} and $s_{T}$ denotes an ordinary Mandelstam---the square of the sum of the external momenta flowing into the tree $T$.  

Note that $\U$ and $\F$ are homogenous polynomials, and that $\F$ is linear in at least one variable provided there is a massless propagator in the graph. Moreover, it is easy to see that the integral (\ref{symanzik_rep}) simplifies considerably when the exponent of the $\U$ polynomial vanishes---that is, when $E\!=\!(L\pl1)D/2$.\footnote{An analogous simplification arises when $E\!=\!L\,D/2$; this case was studied by Brown in ref.\ \cite{Brown:2009ta} (see also refs.\ \cite{Batyrev:1993,Batyrev:1994,Brown:2009rc,Panzer:2016snt})
 with similar conclusions. For our present purposes, however, we note that for dimension $D\!>\!2$, such integrals necessarily involve less-than-\mbox{(the-conjecturally-)}maximal `weight'---as (\ref{symanzik_rep}) would describe a rational integral of dimension $<\!\!L\,D/2$. Moreover, trading $\F$ for $\U$ eliminates all kinematic dependence.} 
We call such integrals {\it marginal}. 

Before moving on, we should describe why we consider marginal integrals especially important for physics: they are capable of having $D$-gon power counting in $D$ dimensions (that is, power counting consistent with one-loop $D$-gon integrals). As such, they represent important irreducible elements of any unitarity basis of loop integrands (and are thus relevant to the scattering amplitudes of all quantum field theories in $D$ dimensions). Another important aspect of marginal integrals---at least in 2 and 4 dimensions---is that these integrals can be shown to have maximal `transcendental weight' (or `weight'). It is not actually known how to make the notion of weight precise beyond polylogarithms (see ref.\ \cite{Broedel:2018qkq} for recent progress in the elliptic case), but here we use it as a proxy for the minimal dimension of algebraic differential form over which it can be expressed as an integral.  We note that, outside of marginal integrals in four dimensions, it can be extremely hard to find a $2L$-fold rational integral representation of a generic Feynman integral. In the cases where it has proven possible, more elaborate methods than described here have been utilized \cite{Bourjaily:2019jrk,Bourjaily:2017bsb,Bourjaily:2018ycu,Bourjaily:2018aeq}.

\newpage\vspace{-18pt}\section{Calabi-Yau Geometry of Loop Integrals}\vspace{-14pt}\label{sec:calabi_yau_geometry_of_feynman_integrals}
In a large and growing number of examples, the (non-polylogarithmic part of the) geometry relevant to individual Feynman integrals has been found to be Calabi-Yau. Examples include the massive $L$-loop banana integrals in two dimensions (see e.g.\ \mbox{refs.\ \cite{Bloch:2014qca,Bloch:2016izu,broadhurstprivate,mirrors_and_sunsets}}) and the $L$-loop massless traintrack integrals \cite{Bourjaily:2017bsb,Bourjaily:2018ycu}, which are known at low loops (and conjectured at high loops) to involve polylogarithms integrated over $(L\mi1)$-dimensional Calabi-Yau manifolds. We are unsure of the extent to which this property holds more generally, but it turns out it is fairly easy to prove that, upon exhausting all rational and polylogarithmic integrations, marginal integrals in any number of dimensions are naturally defined over (possibly singular) Calabi-Yau manifolds. 

The argument is fairly straightforward (and essentially the same as Brown describes in \mbox{ref.\ \cite{Brown:2009ta}} for Feynman integrals with $E\!=\!2\,L$ in $D\!=\!4$). 
 For a marginal Feynman integral, the integrand in (\ref{symanzik_rep}) becomes $1/\F^{D/2}$. Recall that $\F$ is a homogeneous polynomial (of total degree $L\pl1$) in $\mathbb{P}^{E-1}$, and that it is of degree $\leq\!2$ in each homogeneous coordinate. If $\F$ were linear in any variable, we could integrate over it to obtain another rational function, or a polylogarithmic function. If any factor in the denominator were still linear, we could partial-fraction and integrate over it. We may continue in this manner until we have terms whose denominators are irreducibly quadratic or higher in all remaining parameters.

Let us suppose that we have done the above, resulting in a form of the integral which is a polylogarithm integrated over a sum of rational forms whose denominators are each irreducibly quadratic or higher in all $m$ remaining variables. Suppose the forms involve polynomials no worse than quadratic; then any parameter of such an integral can be chosen for one further polylogarithmic integration---but at the cost of introducing a discriminant $Q(x_i)$ of this quadratic with respect to the remaining $(m\mi1)$ variables. It is easy to see that $Q(x_i)$ is homogeneous, with degree $2(m\mi1)$. Thus, the hypersurface $y^2\!=\!Q(x_i)$ defines an algebraic variety in weighted projective space $\mathbb{WP}^{m-1}_{[(m-1):1:\cdots:1]}$, where we assign weight $(m\mi1)$ to $y$ and weight $1$ to the $x_i$. We may think of this as the ``irreducible part'' of the geometry of the Feynman integral; when we refer to the geometry of a given Feynman integral, this is in general what we have in mind.

As the sum of the weights in $\smash{\mathbb{WP}^{m-1}_{[(m-1):1:\cdots:1]}}$ is equal to the degree of $y^2\!=\!Q(x_i)$, we identify the geometry of marginal integrals as Calabi-Yau (as done, for example, in \mbox{ref.\ \cite{Brown:2009ta}}). This is the Calabi-Yau condition mentioned earlier. Since the hypersurface will typically be singular, identifying it with a smooth Calabi-Yau manifold further requires blowing up these singularities \cite{Hubsch:1992nu}. It is possible to carry out this blow-up and identify the resolved manifold as a Calabi-Yau in all cases we are aware of~\cite{Brown:2010bw,OngoingCY}, but we are presently not able to show that this will hold true more generally.


\vspace{-18pt}\section{{\it Exempli Gratia}: Massive Banana Integrals in \texorpdfstring{$D\!=\!2$}{D=2}}
\begin{figure}[t]\vspace{-07pt}\caption{The marginal, $L$-loop banana integral in two dimensions. It has (maximal) rigidity  $(L{-}1)$ {\it iff all legs are massive}.}$\vspace{-0pt}\fwbox{0pt}{\fig{1}{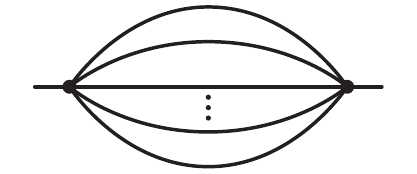}}$\label{banana_figure}\vspace{-10pt}\end{figure}

\vspace{-14pt}\label{sec:banana_integrals}
In two dimensions, the only (tadpole-free) marginal integrals are the so-called `banana' (or `sunrise') graphs depicted in \mbox{Figure \ref{banana_figure}}. It is fairly well known \cite{Bloch:2014qca,Bloch:2016izu,broadhurstprivate} that---in the fully massive case---these integrals have a rigidity of $(L\mi1)$. That is, these integrals may be represented as (one-fold) logarithms integrated over Calabi-Yau manifolds of dimension $(L\mi1)$. For the sake of illustration, let us see how we can understand this fact from the discussion above. 

The Symanzik representation of the $L$-loop banana integral is fairly trivial to derive. For the $\U$ polynomial, there are $1$-forests associated with (not cutting) each edge, and for $\F$, there is a single $2$-forest which cuts every edge. Letting $s$ be the squared-momentum of the external line, we see that we may write $ I\!=\!\int1/\F$ with
\vspace{-2pt}\eq{\begin{split}\U\equivR&\Big(\prod_ix_i\Big)\sum_{i}\frac{1}{x_i}\,,\\\F\equivR&\Big(\prod_ix_i\Big)\Big[s\pl\Big(\sum_{i}\frac{1}{x_i}\Big)\sum_{i}x_im_i^2\Big]\,.\\[-4pt]\end{split}\label{banana_fu_polys}}
Provided $m_i^2\!\neq\!0$ for each $i$, it is easy to see that $\F$ is an irreducible quadratic in each ($\alpha$-parameter) $x_j$. Thus, following the discussion above, we may easily integrate any $x_j$---resulting in a logarithm---but at the cost of introducing the square root associated with the discriminant of $\F$. Suppose that $x_j$ were integrated out; it is not hard to see that the discriminant of $\F$ with respect to $x_j$ is
\vspace{-2pt}\eq{\begin{split}\hspace{-5pt}Q_j\equivR&\Big(\prod_{i\neq j}x_i^2\Big)\Bigg[\!\Big(s\pl m_j^2\pl\Big(\sum_{i\neq j}\frac{1}{x_i}\Big)\sum_{i\neq j}{x_i}m_i^2\Big)^2\\
&\hspace{65pt}\mi4m_j^2\Big(\sum_{i\neq j}\frac{1}{x_i}\Big)\sum_{i\neq j}{x_i}m_i^2\Bigg]\,.\\[-4pt]\end{split}\label{banana_discriminant}}
Integrating over $x_j$ results in an integral over the remaining $L$ parameters of the form $1/\sqrt{Q_j}$ times a logarithm. Notice  that $Q_j$ is homogeneous of degree $2L$---irreducibly quartic in each remaining parameter. Describing the (singular) hypersurface as $y^2\!=\!Q_j$ allows us to embed this as a homogeneous hypersurface in $\mathbb{WP}^L_{[L:1:\cdots:1]}$. As such, the integral is over a Calabi-Yau manifold of dimension $(L\mi1)$. It is worth noting that if even a single mass had vanished, then $\F$ would have been linear in at least one variable, rendering the rigidity of the banana integral strictly less than $(L\mi1)$.

\newpage
\vspace{-18pt}\section{Maximally Rigid, Massless Integrals in \texorpdfstring{$D\!=\!4$}{D=4}}\vspace{-14pt}\label{sec:four_dimensions}
Consider now a finite marginal $L$-loop Feynman integral in four dimensions involving $E\!=\!2(L\pl1)$ massless propagators. Such a Feynman integral is represented in the Symanzik formalism (\ref{symanzik_rep}) by
\vspace{-2pt}\eq{I=\int_{x_i\geq0}\hspace{-17.5pt}\proj{d^{2L+1}x_i}\frac{1}{\F^{2}}\,.\label{four_d_marginal_symanzik}\vspace{-2pt}} 
Because the integral is massless, $\F$ is linear in every variable. As such, we may integrate out any one parameter $x_j$. Writing $\F\!\equivL\F_0^{(j)}\pl x_j\F_{1}^{(j)}$, this results in
\vspace{-2pt}\eq{I=\int_{x_i\geq0}\hspace{-17.5pt}\proj{d^{2L}x_i}\frac{1}{\F_0^{(j)}\F_1^{(j)}}\,.\label{four_d_marginal_symanzik_post_j_integral}\vspace{-2pt}} 
Because $\F$ was linear in each variable, so are each of the factors in the denominator of (\ref{four_d_marginal_symanzik_post_j_integral}); thus, we may always perform another integration resulting in a logarithm, but without introducing any non-rational prefactors. More specifically, suppose we choose to integrate $x_k$; then we may write $\F_i^{(j)}\!\!\equivL\F_{i,0}^{(j,k)}\pl x_k\F_{i,1}^{(j,k)}$ and conclude that
\vspace{-2pt}\eq{\hspace{-30pt}I=\int_{x_i\geq0}\hspace{-17.5pt}\proj{d^{2L-1}x_i}\frac{\displaystyle\log\!\Big(\F_{0,0}^{(j,k)}\F_{1,1}^{(j,k)}\Big)\mi\log\!\Big(\F_{0,1}^{(j,k)}\F_{1,0}^{(j,k)}\Big)}{\F_{0,0}^{(j,k)}\F_{1,1}^{(j,k)}-\F_{0,1}^{(j,k)}\F_{1,0}^{(j,k)}}\,.\hspace{-20pt}\label{four_d_marginal_symanzik_post_jk_integrals}\vspace{-2pt}} 

At this stage we may already conclude that the rigidity of these integrals is $\leq\!2(L\mi1)$: the denominator of (\ref{four_d_marginal_symanzik_post_jk_integrals}) is at most quadratic in each remaining variable, so we can always perform at least one further polylogarithmic integration. In order for an integral to have rigidity exactly $2(L\mi1)$, it must then be the case that, for every choice of the first two integrations, (i) the denominator of (\ref{four_d_marginal_symanzik_post_jk_integrals}) is quadratic in {\it all} remaining variables, and (ii) each quadratic's discriminant is an irreducible quartic or cubic (that is, has no repeated roots) in the remaining $2(L\mi1)$ variables. 

It is worth clarifying the role of condition (ii) above. For an integral satisfying condition (i), performing the third integration results in an integral over the square root of the discriminant of the denominator of (\ref{four_d_marginal_symanzik_post_jk_integrals}). If the polynomial in this square root has repeated roots in any of its variables, we may factor out a perfect square, yielding a linear factor outside of the square root. The remaining polynomial inside the square root is then quadratic, allowing it to be rationalized by a change of variables (see e.g.\ ref.\ \cite{Panzer:2014gra,Besier:2018jen}). This in turn will allow another polylogarithmic integration to be carried out (at least in principle), indicating sub-maximal rigidity.

\begin{figure}[b]\vspace{-10pt}\vspace{-0pt}$\hspace{-10pt}\begin{array}{@{}c@{}}\vspace{-10pt}\begin{array}{@{}c@{}}\fwbox{0pt}{\fig{1}{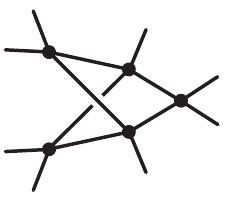}}\\[-4pt](a)\end{array}\hspace{65pt}\begin{array}{@{}c@{}}\fwbox{0pt}{\fig{0.95}{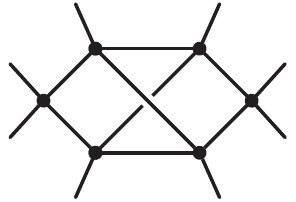}}\\[-4pt](b)\end{array}\hspace{75pt}\begin{array}{@{}c@{}}\fwbox{0pt}{\fig{0.95}{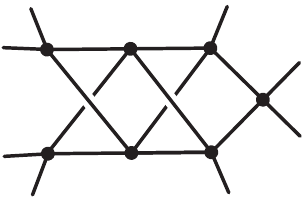}}\\[-4pt](c)\end{array}\\[-5pt]~\\[-6pt]\hspace{-0pt}\begin{array}{@{}c@{}}\fwbox{0pt}{\fig{1}{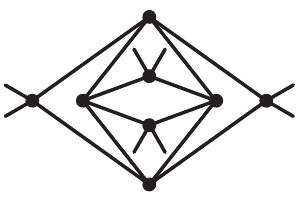}}\\[-2pt](d)\end{array}\hspace{80pt}\begin{array}{@{}c@{}}\fwbox{0pt}{\fig{0.95}{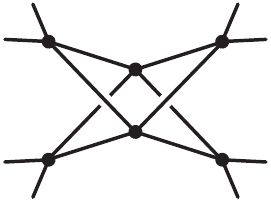}}\\[-2pt](e)\end{array}\\[-10pt]
\end{array}$\caption{A mini-bestiary of (a) two- (b) three- (c) four- and (d) five-loop marginal integrals with maximal rigidity; (e) a marginal integral with {\it sub-maximal} rigidity.}\label{mini_bestiary}\vspace{-3.75pt}\end{figure}

\begin{figure}[t]\vspace{-7pt}\caption{`Tardigrades': massless, finite four-dimensional Feynman integrals with rigidity $2(L{-}1)$ for \emph{even} $L\!\geq\!2$.}\label{tardigrade_figure}\vspace{-8pt}$\vspace{-17pt}\fwbox{0pt}{\fig{1}{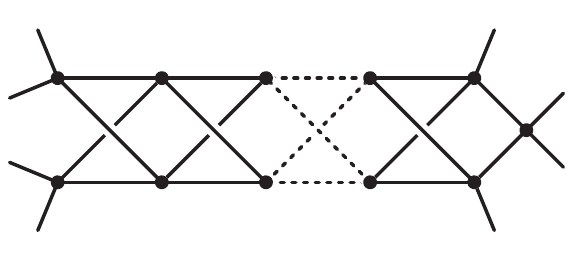}}$\vspace{-12pt}\end{figure}

\begin{figure}[b]\vspace{-26pt}$\vspace{-17pt}\fwbox{0pt}{\fig{1}{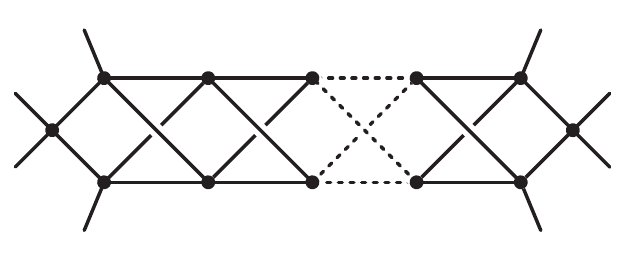}}$\caption{`Paramecia': massless, finite four-dimensional Feynman integrals with rigidity $2(L{-}1)$ for \emph{odd} $L\!\geq\!1$.}\label{paramecia_figure}\vspace{-6pt}\end{figure}

\begin{figure}[t]\vspace{-7pt}\caption{`Amoebas': massless, finite four-dimensional Feynman integrals with rigidity $2(L{-}1)$ for \emph{odd} $L\!\geq\!5$.}\label{amoeba_figure}\vspace{-0.5pt}$\vspace{-18.5pt}\fwbox{0pt}{\fig{1}{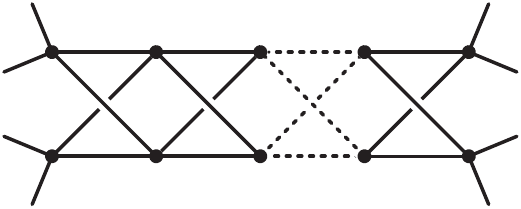}}$\vspace{5.575pt}\end{figure}

Although criteria (i) and (ii) are fairly stringent, it turns out that many marginal, four-dimensional integrals saturate this bound in rigidity. Explicit two- through five-loop finite integrals with maximal rigidity (and one example of a three-loop integral with {\it sub-maximal} rigidity) are shown in \mbox{Figure \ref{mini_bestiary}}. The first four of these are special cases of the all-loop sequences shown in \mbox{Figures \ref{tardigrade_figure}, \ref{paramecia_figure}, and \ref{amoeba_figure}} which we refer to as `tardigrades', `paramecia' and `amoebas', respectively. For these sequences, we have explicitly checked that they saturate the bound on rigidity through eleven loops, and we expect this to hold for all loop orders.

Notice that each of the infinite families of examples depends on a fixed number of external legs. In particular, those shown in \mbox{Figure \ref{amoeba_figure}} can be naturally thought of as `twisted' counterparts to the familiar ladder sequence of integrals \cite{Usyukina:1992jd,Usyukina:1993ch,Broadhurst:1993ib,Broadhurst:2010ds,Drummond:2012bg,Isaev:2003tk,Fleury:2016ykk,Basso:2017jwq}. (In keeping with the spirit of this paper, however, we instead refer to them as amoebas.) Unlike the tardigrades and paramecia, the three-loop amoeba turns out to have non-maximal rigidity; however, it appears to be maximal at all higher loops. It would be worthwhile to investigate whether these diagrams can be re-summed analytically, as was done for the (non-twisted) ladders in \mbox{refs.\  \cite{Usyukina:1992jd,Usyukina:1993ch,Broadhurst:1993ib,Broadhurst:2010ds,Drummond:2012bg}}, and the six-point pentaladders of \mbox{ref.\ \cite{Caron-Huot:2018dsv}}.

\vspace{-18pt}\section{Discussion and Conclusions}\vspace{-14pt}

At any fixed loop order and spacetime dimension (or, equivalently, fixed multiplicity), the scope of Feynman integral complexity is bounded by the finiteness in extent of the relevant loop integrands. In this Letter, we have identified several infinite classes of four-dimensional Feynman integrals---relevant to a wide range of quantum field theories---that involve more complicated geometries at each loop order than all previously known examples. Nevertheless, the extent of relevant geometries is {\it dramatically} more restrictive than even that of Calabi-Yau manifolds. Indeed, there is a real sense in which the fact that these geometries are Calabi-Yau is beside the point: the manifolds relevant to loop integration are {\it much} rarer and {\it much} more special than this.

It is not clear whether the geometry of the (CY) manifold relevant to a given Feynman integral is unique---or if it depends on the order of integrations, for example. In the case which is best understood, the elliptic double box \cite{Bourjaily:2017bsb} (which has rigidity 1), different integration pathways result in different parameterizations of the `same' elliptic curve.\footnote{By this we mean that all pathways give curves with equal moduli.} It would be extremely interesting to know whether or not a similar statement were true for the integrals discussed here: do the hypersurfaces obtained via different integration pathways encode the {\it same} manifold? We expect so; but if this were not the case, it would imply very interesting identities among period integrals over {\it different} manifolds (reminiscent of mirror symmetry, perhaps).

Whether or not the geometry of the hypersurface is uniquely fixed by the graph, our analysis above shows that---at least for the maximally rigid examples discussed---each can be written as a weight-two polylogarithm integrated over some $2(L\mi1)$-dimensional algebraic variety. As such, it may be tempting to view this underlying geometry as fixed, and develop technology for iterated integrals over these geometries (as has been done with considerable success in the case of the elliptic multiple polylogarithms \cite{BrownLevin,Broedel:2014vla,Broedel:2017kkb,Broedel:2017siw,Broedel:2018iwv,Broedel:2018qkq,Broedel:2018qkq}).
 However, it is easy to construct examples\footnote{This can be done by simply adding a propagator to any of the examples discussed here. It is not hard to do this in such a way that the resulting integral includes multiple, distinct copies of the original integral as contact terms.} which involve sums over terms with different geometries (or even different degrees of rigidity).

\paragraph{Wild speculations and conjectures:} Having proven an upper bound on rigidity for the class of marginal Feynman integrals, we conjecture that the same bound holds for all $L$-loop integrals in four dimensions. This conjecture rests on two observations: first, that polylogarithmic integrals are expected to have transcendental weight of at most $2L$ at $L$ loops, and second, that one-loop integrals are always polylogarithmic. Together these suggest that an $L$-loop integral in four dimensions can have only $2L-2$ irreducibly rigid integrations.

It is natural to wonder if a similar statement to our bounded rigidity holds for higher (even) numbers of dimensions. That is, is the rigidity of any $D$-dimensional loop integral bounded by $(L\mi1)D/2$? A possible route to proving such a property might involve Tarasov's dimension-shift relations~\cite{Tarasov:1996br}.

\begin{figure}[t]\vspace{-7pt}\caption{The two most rigid three-loop integrals in the planar limit: (a) the traintrack with rigidity 2 and (b) the `coccolithophore' (or wheel) with rigidity $3$.}\label{cocolithophore}\vspace{-5pt}$\vspace{-17pt}\begin{array}{@{}c@{}}\fwbox{100pt}{\fig{1}{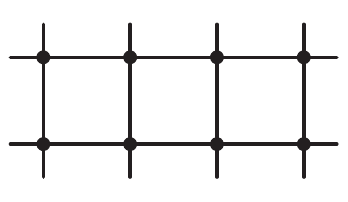}}\\[-4pt](a)\end{array}\begin{array}{@{}c@{}}\fwbox{100pt}{\fig{1}{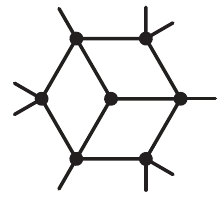}}\\[-4pt](b)\end{array}$\vspace{-0pt}\end{figure}

We have seen that in two and four dimensions, marginal Feynman integrals exist that saturate the rigidity bounds described above. Beyond one loop for $D\!>\!2$, marginality implies non-planarity for integrals with $D$-gon power counting. It is therefore natural to wonder whether stricter rigidity bounds exist for four-dimensional planar integrals. 
In the examples we have considered, an upper bound on the rigidity of a Feynman graph is given by the number of propagators that are `shared' between different loops. 
Thus, we might compare the three-loop traintrack integral~\cite{Bourjaily:2018ycu} with the three-loop `coccolithophore' (a.k.a.\ `wheel') shown in \mbox{Figure \ref{cocolithophore}}. The traintrack has only two shared propagators and rigidity 2, while the coccolithophore has three shared propagators and can be shown to have rigidity~3~\cite{to_appear}. Both graphs have lower rigidity than the three-loop paramecium (\mbox{Figure \ref{mini_bestiary}}b), which saturates our predicted bound having rigidity 4. We expect similar hierarchies to hold at higher loops.

Finally, we should note that the notion of geometric rigidity is rather coarse and in need of considerable refinement and elaboration. For example, an integral over an elliptic curve is clearly simpler than one over a higher-genus surface of the same dimension; and the product of two elliptic curves is clearly less complicated than a generic K3 surface. (Moreover, there are interesting distinctions to be drawn among integrals sharing the same underlying geometry, see e.g.\ \mbox{refs.\ \cite{Blumlein:2018cms,Blumlein:2018jgc}}.) Our notion of rigidity does not allow us to distinguish such cases. An extremely important question going forward would be to develop a better understanding of the (ir)reducibility of the algebraic varieties relevant to Feynman integrals. We hope the examples discussed here may help inspire the development of better tools for understanding (and, ultimately, evaluating) such integrals.

\vspace{6pt}\acknowledgments
We are grateful to Lara Anderson, Spencer Bloch, Claude Duhr, Paul Oehlmann, Erik Panzer, Pierre Vanhove, and Cristian Vergu for helpful discussions, and to Pierre Vanhove for helpful comments on early versions of this draft. 
Finally, we would like to thank Tristan H{\"u}bsch, the original Calabi-Yau beastmaster. 
This project has received funding from the Danish Independent Research Fund under grant number DFF-4002-00037 (MW), the European Union's Horizon 2020 research and innovation program under grant agreement \mbox{No.\ 793151} (MvH), an ERC Starting Grant \mbox{(No.\ 757978)} and a grant from the Villum Fonden (JLB,AJM,MvH,MW).

\vspace{-10pt}
\providecommand{\href}[2]{#2}\begingroup\raggedright\endgroup


\begin{thebibliography}{10}

\vspace{-18pt}
\bibitem{Davydychev:1997wa}
A.~I. Davydychev and R.~Delbourgo, ``{A Geometrical Angle on Feynman
  Integrals},'' \href{http://dx.doi.org/10.1063/1.532513}{{\em J. Math. Phys.}
  {\bf 39} (1998)  4299--4334},
\href{http://arxiv.org/abs/hep-th/9709216}{{ arXiv:hep-th/9709216 [hep-th]}}.

\bibitem{Arkani-Hamed:2017ahv}
N.~Arkani-Hamed and E.~Y. Yuan, ``{One-Loop Integrals from Spherical
  Projections of Planes and Quadrics},''
\href{http://arxiv.org/abs/1712.09991}{{ arXiv:1712.09991 [hep-th]}}.

\bibitem{Beisert:2006ez}
N.~Beisert, B.~Eden, and M.~Staudacher, ``{Transcendentality and Crossing},''
  \href{http://dx.doi.org/10.1088/1742-5468/2007/01/P01021}{{\em J. Stat.
  Mech.} {\bf 0701} (2007)  P01021},
\href{http://arxiv.org/abs/hep-th/0610251}{{ arXiv:hep-th/0610251 [hep-th]}}.

\bibitem{CaronHuot:2011ky}
S.~Caron-Huot, ``{Superconformal Symmetry and Two-Loop Amplitudes in Planar
  $\mathcal{N}\!=\!4$ Super Yang-Mills},''
  \href{http://dx.doi.org/10.1007/JHEP12(2011)066}{{\em JHEP} {\bf 1112} (2011)
   066},
\href{http://arxiv.org/abs/1105.5606}{{ arXiv:1105.5606}}.

\bibitem{Bourjaily:2015bpz}
J.~L. Bourjaily, P.~Heslop, and V.-V. Tran, ``{Perturbation Theory at Eight
  Loops: Novel Structures and the Breakdown of Manifest Conformality in
  $\mathcal{N}\!=\!4$ Supersymmetric Yang-Mills Theory},''
  \href{http://dx.doi.org/10.1103/PhysRevLett.116.191602}{{\em Phys. Rev.
  Lett.} {\bf 116} (2016) no. 19, 191602},
\href{http://arxiv.org/abs/1512.07912}{{ arXiv:1512.07912 [hep-th]}}.

\bibitem{Bourjaily:2016evz}
J.~L. Bourjaily, P.~Heslop, and V.-V. Tran, ``{Amplitudes and Correlators to
  Ten Loops Using Simple, Graphical Bootstraps},''
  \href{http://dx.doi.org/10.1007/JHEP11(2016)125}{{\em JHEP} {\bf 11} (2016)
  125},
\href{http://arxiv.org/abs/1609.00007}{{ arXiv:1609.00007}}.

\bibitem{Henn:2016jdu}
J.~M. Henn and B.~Mistlberger, ``{Four-Gluon Scattering at Three Loops,
  Infrared Structure, and the Regge Limit},''
  \href{http://dx.doi.org/10.1103/PhysRevLett.117.171601}{{\em Phys. Rev.
  Lett.} {\bf 117} (2016) no. 17, 171601},
\href{http://arxiv.org/abs/1608.00850}{{ arXiv:1608.00850 [hep-th]}}.

\bibitem{DelDuca:2016lad}
V.~Del~Duca, S.~Druc, J.~Drummond, C.~Duhr, F.~Dulat, R.~Marzucca,
  G.~Papathanasiou, and B.~Verbeek, ``{Multi-Regge Kinematics and the Moduli
  Space of Riemann Spheres with Marked Points},''
  \href{http://dx.doi.org/10.1007/JHEP08(2016)152}{{\em JHEP} {\bf 08} (2016)
  152},
\href{http://arxiv.org/abs/1606.08807}{{ arXiv:1606.08807 [hep-th]}}.

\bibitem{Caron-Huot:2016owq}
S.~Caron-Huot, L.~J. Dixon, A.~McLeod, and M.~von Hippel, ``{Bootstrapping a
  Five-Loop Amplitude Using Steinmann Relations},''
  \href{http://dx.doi.org/10.1103/PhysRevLett.117.241601}{{\em Phys. Rev.
  Lett.} {\bf 117} (2016) no. 24, 241601},
\href{http://arxiv.org/abs/1609.00669}{{ arXiv:1609.00669 [hep-th]}}.

\bibitem{Dixon:2016nkn}
L.~J. Dixon, J.~Drummond, T.~Harrington, A.~J. McLeod, G.~Papathanasiou, and
  M.~Spradlin, ``{Heptagons from the Steinmann Cluster Bootstrap},''
  \href{http://dx.doi.org/10.1007/JHEP02(2017)137}{{\em JHEP} {\bf 02} (2017)
  137},
\href{http://arxiv.org/abs/1612.08976}{{ arXiv:1612.08976 [hep-th]}}.

\bibitem{Almelid:2017qju}
{\O}.~Almelid, C.~Duhr, E.~Gardi, A.~McLeod, and C.~D. White, ``{Bootstrapping
  the QCD Soft Anomalous Dimension},''
  \href{http://dx.doi.org/10.1007/JHEP09(2017)073}{{\em JHEP} {\bf 09} (2017)
  073},
\href{http://arxiv.org/abs/1706.10162}{{ arXiv:1706.10162}}.

\bibitem{Broadhurst:1993mw}
D.~J. Broadhurst, J.~Fleischer, and O.~V. Tarasov, ``{Two-Loop Two-Point
  Functions with Masses: Asymptotic Expansions and Taylor Series, in Any
  Dimension},'' \href{http://dx.doi.org/10.1007/BF01474625}{{\em Z. Phys.} {\bf
  C60} (1993)  287--302},
\href{http://arxiv.org/abs/hep-ph/9304303}{{ arXiv:hep-ph/9304303 [hep-ph]}}.

\bibitem{Bloch:2013tra}
S.~Bloch and P.~Vanhove, ``{The Elliptic Dilogarithm for the Sunset Graph},''
  \href{http://dx.doi.org/doi:10.1016/j.jnt.2014.09.032}{{\em J. Number Theory}
  {\bf 148} (2015)  328--364},
\href{http://arxiv.org/abs/1309.5865}{{ arXiv:1309.5865 [hep-th]}}.

\bibitem{Bloch:2016izu}
S.~Bloch, M.~Kerr, and P.~Vanhove, ``{Local Mirror Symmetry and the Sunset
  Feynman Integral},''
  \href{http://dx.doi.org/10.4310/ATMP.2017.v21.n6.a1}{{\em Adv. Theor. Math.
  Phys.} {\bf 21} (2017)  1373--1453},
\href{http://arxiv.org/abs/1601.08181}{{ arXiv:1601.08181 [hep-th]}}.

\bibitem{Bloch:2014qca}
S.~Bloch, M.~Kerr, and P.~Vanhove, ``{A Feynman Integral via Higher Normal
  Functions},'' \href{http://dx.doi.org/10.1112/S0010437X15007472}{{\em Compos.
  Math.} {\bf 151} (2015) no. 12, 2329--2375},
\href{http://arxiv.org/abs/1406.2664}{{ arXiv:1406.2664}}.

\bibitem{Broadhurst:2016myo}
D.~Broadhurst, ``{Feynman Integrals, $L$-Series and Kloosterman Moments},''
  \href{http://dx.doi.org/10.4310/CNTP.2016.v10.n3.a3}{{\em Commun. Num. Theor.
  Phys.} {\bf 10} (2016)  527--569},
\href{http://arxiv.org/abs/1604.03057}{{ arXiv:1604.03057 [physics.gen-ph]}}.

\bibitem{Bogner:2017vim}
C.~Bogner, A.~Schweitzer, and S.~Weinzierl, ``{Analytic Continuation and
  Numerical Evaluation of the Kite Integral and the Equal Mass Sunrise
  Integral},'' \href{http://dx.doi.org/10.1016/j.nuclphysb.2017.07.008}{{\em
  Nucl. Phys.} {\bf B922} (2017)  528--550},
\href{http://arxiv.org/abs/1705.08952}{{ arXiv:1705.08952 [hep-ph]}}.

\bibitem{CaronHuot:2012ab}
S.~Caron-Huot and K.~J. Larsen, ``{Uniqueness of Two-Loop Master Contours},''
  \href{http://dx.doi.org/10.1007/JHEP10(2012)026}{{\em JHEP} {\bf 1210} (2012)
   026},
\href{http://arxiv.org/abs/1205.0801}{{ arXiv:1205.0801 [hep-ph]}}.

\bibitem{ArkaniHamed:2012nw}
N.~Arkani-Hamed, J.~L. Bourjaily, F.~Cachazo, A.~B. Goncharov, A.~Postnikov,
  and J.~Trnka, ``{Scattering Amplitudes and the Positive Grassmannian},''
\href{http://arxiv.org/abs/1212.5605}{{ arXiv:1212.5605 [hep-th]}}.

\bibitem{Bourjaily:2015jna}
J.~L. Bourjaily and J.~Trnka, ``{Local Integrand Representations of All
  Two-Loop Amplitudes in Planar SYM},''
  \href{http://dx.doi.org/10.1007/JHEP08(2015)119}{{\em JHEP} {\bf 08} (2015)
  119},
\href{http://arxiv.org/abs/1505.05886}{{ arXiv:1505.05886 [hep-th]}}.

\bibitem{Bourjaily:2017wjl}
J.~L. Bourjaily, E.~Herrmann, and J.~Trnka, ``{Prescriptive Unitarity},''
  \href{http://dx.doi.org/10.1007/JHEP06(2017)059}{{\em JHEP} {\bf 06} (2017)
  059},
\href{http://arxiv.org/abs/1704.05460}{{ arXiv:1704.05460 [hep-th]}}.

\bibitem{Bourjaily:2017bsb}
J.~L. Bourjaily, A.~J. McLeod, M.~Spradlin, M.~von Hippel, and M.~Wilhelm,
  ``{Elliptic Double-Box Integrals: Massless Scattering Amplitudes beyond
  Polylogarithms},''
  \href{http://dx.doi.org/10.1103/PhysRevLett.120.121603}{{\em Phys. Rev.
  Lett.} {\bf 120} (2018) no. 12, 121603},
\href{http://arxiv.org/abs/1712.02785}{{ arXiv:1712.02785 [hep-th]}}.

\bibitem{Bourjaily:2018ycu}
J.~L. Bourjaily, Y.-H. He, A.~J. Mcleod, M.~Von~Hippel, and M.~Wilhelm,
  ``{Traintracks Through Calabi-Yaus: Amplitudes Beyond Elliptic
  Polylogarithms},''
  \href{http://dx.doi.org/10.1103/PhysRevLett.121.071603}{{\em Phys. Rev.
  Lett.} {\bf 121} (2018) no. 7, 071603},
\href{http://arxiv.org/abs/1805.09326}{{ arXiv:1805.09326}}.

\bibitem{Caffo:1998du}
M.~Caffo, H.~Czyz, S.~Laporta, and E.~Remiddi, ``{The Master Differential
  Equations for the Two-Loop Sunrise Selfmass Amplitudes},'' {\em Nuovo Cim.}
  {\bf A111} (1998)  365--389,
\href{http://arxiv.org/abs/hep-th/9805118}{{ arXiv:hep-th/9805118 [hep-th]}}.

\bibitem{Laporta:2004rb}
S.~Laporta and E.~Remiddi, ``{Analytic Treatment of the Two-Loop Equal Mass
  Sunrise Graph},''
  \href{http://dx.doi.org/10.1016/j.nuclphysb.2004.10.044}{{\em Nucl. Phys.}
  {\bf B704} (2005)  349--386},
\href{http://arxiv.org/abs/hep-ph/0406160}{{ arXiv:hep-ph/0406160 [hep-ph]}}.

\bibitem{Kniehl:2005bc}
B.~A. Kniehl, A.~V. Kotikov, A.~Onishchenko, and O.~Veretin, ``{Two-Loop Sunset
  Diagrams with Three Massive Lines},''
  \href{http://dx.doi.org/10.1016/j.nuclphysb.2006.01.013}{{\em Nucl. Phys.}
  {\bf B738} (2006)  306--316},
\href{http://arxiv.org/abs/hep-ph/0510235}{{ arXiv:hep-ph/0510235 [hep-ph]}}.

\bibitem{Groote:2005ay}
S.~Groote, J.~G. Korner, and A.~A. Pivovarov, ``{On the Evaluation of a Certain
  Class of Feynman Diagrams in $x$-Space: Sunrise-Type Topologies at Any Loop
  Order},'' \href{http://dx.doi.org/10.1016/j.aop.2006.11.001}{{\em Annals
  Phys.} {\bf 322} (2007)  2374--2445},
\href{http://arxiv.org/abs/hep-ph/0506286}{{ arXiv:hep-ph/0506286 [hep-ph]}}.

\bibitem{Groote:2012pa}
S.~Groote, J.~G. Korner, and A.~A. Pivovarov, ``{A Numerical Test of
  Differential Equations for One- and Two-Loop sunrise Diagrams using
  Configuration Space Techniques},''
  \href{http://dx.doi.org/10.1140/epjc/s10052-012-2085-z}{{\em Eur. Phys. J.}
  {\bf C72} (2012)  2085},
\href{http://arxiv.org/abs/1204.0694}{{ arXiv:1204.0694 [hep-ph]}}.

\bibitem{Bailey:2008ib}
D.~H. Bailey, J.~M. Borwein, D.~Broadhurst, and M.~L. Glasser, ``{Elliptic
  Integral Evaluations of Bessel Moments},''
  \href{http://dx.doi.org/10.1088/1751-8113/41/20/205203}{{\em J. Phys.} {\bf
  A41} (2008)  205203},
\href{http://arxiv.org/abs/0801.0891}{{ arXiv:0801.0891 [hep-th]}}.

\bibitem{MullerStach:2011ru}
S.~M{\"u}ller-Stach, S.~Weinzierl, and R.~Zayadeh, ``{A Second-Order
  Differential Equation for the Two-Loop Sunrise Graph with Arbitrary
  Masses},'' \href{http://dx.doi.org/10.4310/CNTP.2012.v6.n1.a5}{{\em Commun.
  Num. Theor. Phys.} {\bf 6} (2012)  203--222},
\href{http://arxiv.org/abs/1112.4360}{{ arXiv:1112.4360}}.

\bibitem{Adams:2013kgc}
L.~Adams, C.~Bogner, and S.~Weinzierl, ``{The Two-Loop Sunrise Graph with
  Arbitrary Masses},'' \href{http://dx.doi.org/10.1063/1.4804996}{{\em J. Math.
  Phys.} {\bf 54} (2013)  052303},
\href{http://arxiv.org/abs/1302.7004}{{ arXiv:1302.7004 [hep-ph]}}.

\bibitem{Remiddi:2013joa}
E.~Remiddi and L.~Tancredi, ``{Schouten Identities for Feynman Graph
  Amplitudes: the Master Integrals for the Two-Loop Massive Sunrise Graph},''
  \href{http://dx.doi.org/10.1016/j.nuclphysb.2014.01.009}{{\em Nucl. Phys.}
  {\bf B880} (2014)  343--377},
\href{http://arxiv.org/abs/1311.3342}{{ arXiv:1311.3342 [hep-ph]}}.

\bibitem{Adams:2014vja}
L.~Adams, C.~Bogner, and S.~Weinzierl, ``{The Two-Loop Sunrise Graph in Two
  Space-Time Dimensions with Arbitrary Masses in Terms of Elliptic
  Dilogarithms},'' \href{http://dx.doi.org/10.1063/1.4896563}{{\em J. Math.
  Phys.} {\bf 55} (2014) no. 10, 102301},
\href{http://arxiv.org/abs/1405.5640}{{ arXiv:1405.5640}}.

\bibitem{Adams:2015ydq}
L.~Adams, C.~Bogner, and S.~Weinzierl, ``{The Iterated Structure of the
  All-Order Result for the Two-Loop Sunrise Integral},''
  \href{http://dx.doi.org/10.1063/1.4944722}{{\em J. Math. Phys.} {\bf 57}
  (2016) no. 3, 032304},
\href{http://arxiv.org/abs/1512.05630}{{ arXiv:1512.05630 [hep-ph]}}.

\bibitem{Adams:2017ejb}
L.~Adams and S.~Weinzierl, ``{Feynman Integrals and Iterated Integrals of
  Modular Forms},'' \href{http://dx.doi.org/10.4310/CNTP.2018.v12.n2.a1}{{\em
  Commun. Num. Theor. Phys.} {\bf 12} (2018)  193--251},
\href{http://arxiv.org/abs/1704.08895}{{ arXiv:1704.08895}}.

\bibitem{Remiddi:2016gno}
E.~Remiddi and L.~Tancredi, ``{Differential Equations and Dispersion Relations
  for Feynman Amplitudes: the Two-Loop Massive Sunrise and the Kite
  Integral},'' \href{http://dx.doi.org/10.1016/j.nuclphysb.2016.04.013}{{\em
  Nucl. Phys.} {\bf B907} (2016)  400--444},
\href{http://arxiv.org/abs/1602.01481}{{ arXiv:1602.01481}}.

\bibitem{Adams:2016xah}
L.~Adams, C.~Bogner, A.~Schweitzer, and S.~Weinzierl, ``{The Kite Integral to
  All Orders in Terms of Elliptic Polylogarithms},''
  \href{http://dx.doi.org/10.1063/1.4969060}{{\em J. Math. Phys.} {\bf 57}
  (2016) no. 12, 122302},
\href{http://arxiv.org/abs/1607.01571}{{ arXiv:1607.01571 [hep-ph]}}.

\bibitem{Nandan:2013ip}
D.~Nandan, M.~F. Paulos, M.~Spradlin, and A.~Volovich, ``{Star Integrals,
  Convolutions and Simplices},''
  \href{http://dx.doi.org/10.1007/JHEP05(2013)105}{{\em JHEP} {\bf 05} (2013)
  105},
\href{http://arxiv.org/abs/1301.2500}{{ arXiv:1301.2500 [hep-th]}}.

\bibitem{Paulos:2012nu}
M.~F. Paulos, M.~Spradlin, and A.~Volovich, ``{Mellin Amplitudes for Dual
  Conformal Integrals},'' \href{http://dx.doi.org/10.1007/JHEP08(2012)072}{{\em
  JHEP} {\bf 08} (2012)  072},
\href{http://arxiv.org/abs/1203.6362}{{ arXiv:1203.6362 [hep-th]}}.

\bibitem{broadhurstprivate}
D.~Broadhurst \!\!, private communication.

\bibitem{mirrors_and_sunsets}
C.~F. Doran, A.~Y. Novoseltsev, and P.~Vanhove, ``{Mirror Symmetry and Sunset
  Feynman Integrals}.'' {\it To appear}.

\bibitem{Smirnov:2004ym}
V.~A. Smirnov, ``{Evaluating Feynman Integrals},''
\href{http://dx.doi.org/10.1007/b95498}{{\em Springer Tracts Mod. Phys.} {\bf
  211} (2004)  1--244}.

\bibitem{Brown:2009ta}
F.~C.~S. Brown, ``{On the Periods of Some Feynman Integrals},''
\href{http://arxiv.org/abs/0910.0114}{{ arXiv:0910.0114 [math.AG]}}.

\bibitem{Batyrev:1993}
V.~V. Batyrev, ``{Variations of the Mixed Hodge Structure of Affine
  Hypersurfaces in Algebraic Tori},''
  \href{http://dx.doi.org/10.1215/S0012-7094-93-06917-7}{{\em Duke Mathematical
  Journal} {\bf 69} (1993) no. 2, 349--409}.

\bibitem{Batyrev:1994}
V.~V. Batyrev and D.~A. Cox, ``{On the Hodge Structure of Projective
  Hypersurfaces in Toric Varieties},''
  \href{http://dx.doi.org/10.1215/S0012-7094-94-07509-1}{{\em Duke Mathematical
  Journal} {\bf 75} (1994) no. 2, 293--338}.

\bibitem{Brown:2009rc}
F.~Brown and K.~Yeats, ``{Spanning Forest Polynomials and the Transcendental
  Weight of Feynman Graphs},''
  \href{http://dx.doi.org/10.1007/s00220-010-1145-1}{{\em Commun. Math. Phys.}
  {\bf 301} (2011)  357--382},
\href{http://arxiv.org/abs/0910.5429}{{ arXiv:0910.5429 [math-ph]}}.

\bibitem{Panzer:2016snt}
E.~Panzer and O.~Schnetz, ``{The Galois Coaction on $\varphi^4$ Periods},''
  \href{http://dx.doi.org/10.4310/CNTP.2017.v11.n3.a3}{{\em Commun. Num. Theor.
  Phys.} {\bf 11} (2017)  657--705},
\href{http://arxiv.org/abs/1603.04289}{{ arXiv:1603.04289 [hep-th]}}.

\bibitem{Broedel:2018qkq}
J.~Br\"{o}del, C.~Duhr, F.~Dulat, B.~Penante, and L.~Tancredi, ``{Elliptic
  Feynman Integrals and Pure Functions},''
\href{http://arxiv.org/abs/1809.10698}{{ arXiv:1809.10698 [hep-th]}}.

\bibitem{Bourjaily:2019jrk} 
  J.~L.~Bourjaily, F.~Dulat and E.~Panzer,
  ``{Manifestly Dual-Conformal Loop Integration},''
  \href{http://arxiv.org/abs/1901.02887}{{ arXiv:1901.02887 [hep-th]}}.
  
\bibitem{Bourjaily:2018aeq}
J.~L. Bourjaily, A.~J. McLeod, M.~von Hippel, and M.~Wilhelm, ``{Rationalizing
  Loop Integration},'' \href{http://dx.doi.org/10.1007/JHEP08(2018)184}{{\em
  JHEP} {\bf 08} (2018)  184},
\href{http://arxiv.org/abs/1805.10281}{{ arXiv:1805.10281 [hep-th]}}.

\bibitem{Brown:2010bw} 
  F.~Brown and O.~Schnetz,
  ``{A K3 in $\phi^4$},'' \href{http://dx.doi.org/10.1215/00127094-1644201}{{\em
  Duke Math. J.} {\bf 161} (2012)  no. 10, 1817},
\href{http://arxiv.org/abs/1006.4064}{{ arXiv:1006.4064 [math.AG]}}.


\bibitem{OngoingCY}
J.~L. Bourjaily, P.~Candelas, M.~Elmi, A.~J. McLeod, S.~Schafer-Nameki, M.~von Hippel, Y.~Wang, and M.~Wilhelm. {\it Ongoing work}.


\bibitem{Hubsch:1992nu}
T.~Hubsch, \href{http://dx.doi.org/10.1142/1410}{{\em {Calabi-Yau Manifolds: A
  Bestiary for Physicists}}}.
\newblock World Scientific, Singapore,
1994.
\newblock

\bibitem{Panzer:2014gra}
E.~Panzer, ``{On Hyperlogarithms and Feynman Integrals with Divergences and
  Many Scales},'' \href{http://dx.doi.org/10.1007/JHEP03(2014)071}{{\em JHEP}
  {\bf 03} (2014)  071},
\href{http://arxiv.org/abs/1401.4361}{{ arXiv:1401.4361 [hep-th]}}.

\bibitem{Besier:2018jen}
M.~Besier, D.~van Straten, and S.~Weinzierl, ``{Rationalizing Roots: an
  Algorithmic Approach},''
\href{http://arxiv.org/abs/1809.10983}{{ arXiv:1809.10983 [hep-th]}}.

\bibitem{Usyukina:1992jd}
N.~I. Usyukina and A.~I. Davydychev, ``{An Approach to the Evaluation of Three
  and Four Point Ladder Diagrams},''
\href{http://dx.doi.org/10.1016/0370-2693(93)91834-A}{{\em Phys. Lett.} {\bf
  B298} (1993)  363--370}.

\bibitem{Usyukina:1993ch}
N.~I. Usyukina and A.~I. Davydychev, ``{Exact Results for Three and Four Point
  Ladder Diagrams with an Arbitrary Number of Rungs},''
\href{http://dx.doi.org/10.1016/0370-2693(93)91118-7}{{\em Phys. Lett.} {\bf
  B305} (1993)  136--143}.

\bibitem{Broadhurst:1993ib}
D.~J. Broadhurst, ``{Summation of an Infinite Series of Ladder Diagrams},''
\href{http://dx.doi.org/10.1016/0370-2693(93)90202-S}{{\em Phys. Lett.} {\bf
  B307} (1993)  132--139}.

\bibitem{Broadhurst:2010ds}
D.~J. Broadhurst and A.~I. Davydychev, ``{Exponential Suppression with Four
  Legs and an Infinity of Loops},''
  \href{http://dx.doi.org/10.1016/j.nuclphysbps.2010.09.014}{{\em Nucl. Phys.
  Proc. Suppl.} {\bf 205-206} (2010)  326--330},
\href{http://arxiv.org/abs/1007.0237}{{ arXiv:1007.0237 [hep-th]}}.

\bibitem{Drummond:2012bg}
J.~M. Drummond, ``{Generalised Ladders and Single-Valued Polylogarithms},''
  \href{http://dx.doi.org/10.1007/JHEP02(2013)092}{{\em JHEP} {\bf 02} (2013)
  092},
\href{http://arxiv.org/abs/1207.3824}{{ arXiv:1207.3824 [hep-th]}}.

\bibitem{Isaev:2003tk}
A.~P. Isaev, ``{Multiloop Feynman Integrals and Conformal Quantum Mechanics},''
  \href{http://dx.doi.org/10.1016/S0550-3213(03)00393-6}{{\em Nucl. Phys.} {\bf
  B662} (2003)  461--475},
\href{http://arxiv.org/abs/hep-th/0303056}{{ arXiv:hep-th/0303056 [hep-th]}}.

\bibitem{Fleury:2016ykk}
T.~Fleury and S.~Komatsu, ``{Hexagonalization of Correlation Functions},''
  \href{http://dx.doi.org/10.1007/JHEP01(2017)130}{{\em JHEP} {\bf 01} (2017)
  130},
\href{http://arxiv.org/abs/1611.05577}{{ arXiv:1611.05577 [hep-th]}}.

\bibitem{Basso:2017jwq}
B.~Basso and L.~J. Dixon, ``{Gluing Ladder Feynman Diagrams into Fishnets},''
  \href{http://dx.doi.org/10.1103/PhysRevLett.119.071601}{{\em Phys. Rev.
  Lett.} {\bf 119} (2017) no. 7, 071601},
\href{http://arxiv.org/abs/1705.03545}{{ arXiv:1705.03545 [hep-th]}}.

\bibitem{Caron-Huot:2018dsv}
S.~Caron-Huot, L.~J. Dixon, M.~von Hippel, A.~J. McLeod, and G.~Papathanasiou,
  ``{The Double Pentaladder Integral to All Orders},''
  \href{http://dx.doi.org/10.1007/JHEP07(2018)170}{{\em JHEP} {\bf 07} (2018)
  170},
\href{http://arxiv.org/abs/1806.01361}{{ arXiv:1806.01361 [hep-th]}}.

\bibitem{BrownLevin}
F.~Brown and A.~Levin, ``{Multiple Elliptic Polylogarithms},''
  \href{http://arxiv.org/abs/1110.6917}{{ arXiv:1110.6917}}.

\bibitem{Broedel:2014vla}
J.~Br\"{o}del, C.~R. Mafra, N.~Matthes, and O.~Schlotterer, ``{Elliptic
  Multiple Zeta Values and One-Loop Superstring Amplitudes},''
  \href{http://dx.doi.org/10.1007/JHEP07(2015)112}{{\em JHEP} {\bf 07} (2015)
  112},
\href{http://arxiv.org/abs/1412.5535}{{ arXiv:1412.5535 [hep-th]}}.

\bibitem{Broedel:2017kkb}
J.~Br{\"o}del, C.~Duhr, F.~Dulat, and L.~Tancredi, ``{Elliptic polylogarithms
  and iterated integrals on elliptic curves. Part I: general formalism},''
  \href{http://dx.doi.org/10.1007/JHEP05(2018)093}{{\em JHEP} {\bf 05} (2018)
  093},
\href{http://arxiv.org/abs/1712.07089}{{ arXiv:1712.07089 [hep-th]}}.

\bibitem{Broedel:2017siw}
J.~Br\"{o}del, C.~Duhr, F.~Dulat, and L.~Tancredi, ``{Elliptic Polylogarithms
  and Iterated Integrals on Elliptic Curves II: an Application to the Sunrise
  Integral},'' \href{http://dx.doi.org/10.1103/PhysRevD.97.116009}{{\em Phys.
  Rev.} {\bf D97} (2018) no. 11, 116009},
\href{http://arxiv.org/abs/1712.07095}{{ arXiv:1712.07095 [hep-ph]}}.

\bibitem{Broedel:2018iwv}
J.~Br\"{o}del, C.~Duhr, F.~Dulat, B.~Penante, and L.~Tancredi, ``{Elliptic
  symbol calculus: from elliptic polylogarithms to iterated integrals of
  Eisenstein series},'' \href{http://dx.doi.org/10.1007/JHEP08(2018)014}{{\em
  JHEP} {\bf 08} (2018)  014},
\href{http://arxiv.org/abs/1803.10256}{{ arXiv:1803.10256 [hep-th]}}.

\bibitem{Tarasov:1996br} 
  O.~V.~Tarasov,
  ``{Connection between Feynman integrals having different values of the space-time dimension},'' \href{http://dx.doi.org/10.1103/PhysRevD.54.6479}{{\em Phys.
  Rev.} {\bf D54} (1996) 6479-6490},
\href{http://arxiv.org/abs/hep-th/9606018}{{ arXiv:hep-th/9606018}}.

\bibitem{to_appear}
J.~L. Bourjaily, A.~J. McLeod, M.~von Hippel, and M.~Wilhelm. {\it In prep}.

\bibitem{Blumlein:2018cms}
J.~Bl{\"u}mlein and C.~Schneider, ``{Analytic Computing Methods for Precision
  Calculations in Quantum Field Theory},''
  \href{http://dx.doi.org/10.1142/S0217751X18300156}{{\em Int. J. Mod. Phys.}
  {\bf A33} (2018) no. 17, 1830015},
\href{http://arxiv.org/abs/1809.02889}{{ arXiv:1809.02889 [hep-ph]}}.

\bibitem{Blumlein:2018jgc}
J.~Bl{\"u}mlein, ``{Iterative Non-iterative Integrals in Quantum Field
  Theory},'' in {\em {KMPB Conference: Elliptic Integrals, Elliptic Functions
  and Modular Forms in Quantum Field Theory Zeuthen, Germany, October 23-26,
  2017}}.
\newblock 2018.
\newblock
\href{http://arxiv.org/abs/1808.08128}{{ arXiv:1808.08128 [hep-th]}}.
\newblock

\end{thebibliography}
\end{document}